\documentclass[aps,prb,amsmath,amssymb,unsortedaddress]{revtex4}
\usepackage{graphicx}
\usepackage{natbib}

\begin{document}
\title{THE PREFERRED SYSTEM OF REFERENCE RELOADED}
\author{Israel Perez}
\email{cooguion@yahoo.com}
\affiliation{{\it Department of Physics and Engineering Physics, University of Saskatchewan, Saskatoon, SK, Canada}}

\author{The present topic won a prize awarded on November 2012 by the Foundational Question Institute (FQXi)  in an essay competition: Which of our basic physical assumptions are wrong?}


\begin{abstract}
According to Karl Popper assumptions are statements used to construct theories. During the construction of a theory whether the assumptions are either true or false turn out to be irrelevant in view of the fact that, actually, they gain their scientific value when the deductions derived from them suffice to explain observations. Science is enriched with assumptions of all kinds and physics is not exempted. Beyond doubt, some assumptions have been greatly beneficial for physics. They are usually embraced based on the kind of problems expected to be solved in a given moment of a science. Some have been quite useful and some others are discarded in a given moment and reconsidered in a later one. An illustrative example of this is the conception of light; first, according to Newton, as particle; then, according to Huygens, as wave; and then, again, according to Einstein, as particle. Likewise, once, according to Newton, a preferred system of reference (PSR) was assumed; then, according to Einstein, rejected; and then, here the assumption is reconsidered. It is claimed that the assumption that there is no PSR can be fundamentally wrong.
\end{abstract}

\maketitle
\section{Introduction}
One of the main objectives of present day physics is to formulate the so-called theory of everything (TOE), a unifying theory that will be capable of describing physical phenomena at all spatial and energy scales. For the past fifty years, legions of physicists have worked relentlessly on this problem without arriving at satisfactory results. This lack of success tells us two important things: first, that the problem is much more complex than originally thought; and, second, that we may have arrived at a dead end. Indeed, many researchers are realizing that theoretical physics is falling into a deep crisis. Such crisis is mirrored in an ever increasing number of publications proposing bold alternatives and yet no clear answers have been found. Actually, as more experimental evidence piles up, the puzzle aggravates. When we found ourselves in such situation, the most natural way to proceed is to revise the foundations and discard that that is blinding our sight. Here we shall review one of the most prominent principles in the history of physics that, if seriously reconsidered, can help to heal the current problems in physics, namely, the existence of a preferred system of reference (PSR).

\section{Epistemological background} 

Before we move on to our central topic, it would be appropriate to discuss the epistemology behind scientific theories. In this section we shall inform the reader that certain kind of `scientific hypothesis' are natural components of physical theories and that they are legitimate insofar as they reinforce the theory. In doing this, we shall present a series of arguments that would be pivotal to understand why the PSR can still play a major role in the future of physics.

\subsection{Assumptions: True, false or useful?}

For the benefit of this work, we shall see as synonyms the words: postulate, axiom and principle. By these we understand a statement or proposition that serves as the foundation for a chain of reasoning in the construction of a theory. Generally, it is understood that this kind of propositions are true because they are so self-evident that no proof is demanded to demonstrate their veracity. For instance, the proposition `{\it A straight line segment can be drawn joining any two points}' is generally accepted as true. However, when giving a deeper thought, some propositions turn out to be uncertain. To illustrate this point, let us consider the following assertion: {\it Space and time are continuous}. If we now ask: is this statement true or false? Needless to say, the reply would be a shrug. In the face of the lack of certainty, the theorist can still proceed and argue that during the conception of a theory whether the statement is false or true turns out to be irrelevant provided that the predictions derived from such assumption reproduce the experimental evidence at hand. In other words, the assumption will gain its scientific value not from the preliminary judgement of the statement but from the experimental verification of the predictions which are derived from it. Only then, the theorist can vigorously contend that the assumption is not only true but also scientific. Thus, despite that the truthfulness or falsehood of a statement can be debatable, a theorist can {\it presume, for practical purposes}, that the proposition has the potential to be really true. 

Theories, on the other hand, can also rest on the basis of `false' statements as long as these are helpful to strengthen the theory. Examples of this sort can be found anywhere in physics. One typical case is the possibility of reversible processes in thermodynamics. After all, we all are aware that this notion was invented having in mind that, in practice, all processes are irreversible. Despite this, the principle has been highly beneficial for this branch of physics. Other commonly used `false' assumptions are: point particles or rigid bodies. Therefore, at the end, for the theorist what is crucial is not the truthfulness or falsehood of the assumptions but their {\it usefulness} in solving particular problems.

 \subsection{Physical theories}
In a wide sense, we understand by physical theory a rational and logical construct composed of principles, concepts and definitions aim at explaining experimental observations. Theories are formulated by seeking {\it correlations and symmetries} among the observations. The summary of this search is then etched in the so-called laws of nature.\cite{mittelstaedt} These are a set of statements written in mathematical language where the information of physical phenomena has been, so to speak, codified. According to Max Tegmark\cite{tegmark1,tegmark5} theories have two components: mathematical structures (equations) and baggage. Baggage are words that explain how the theories are connected to what we humans observe and intuitively understand, i.e, ontologies, `physical' concepts and notions. And since physics is mainly written in mathematical language, Tegmark goes beyond and asserts that the universe is actually a complex mathematical structure. He remarks that as the theory becomes more fundamental, the level of abstraction increases up to the point that only mathematics would be capable of describing the universe and, therefore, the baggage would be gradually replaced by mathematics. According to him the TOE will have no baggage at all. At first sight, this position seems to be extremist but a deeper reflexion shows us that it may not be the case. To grasp the significance of this, first, one should ask what a mathematical structure is. The minimalistic view is that a mathematical structure is no other thing that a set of abstract objects connected by logical relations and operations. For our purposes, this definition suffices since the task of physics is to seek for `physical' correlations. From this standpoint, one is then allowed to assert that the description of the universe can be reduced to a set of logical relations, i.e., physical laws. If we agree, this means that what can be said of the universe in terms of baggage can also be said with mathematics. Mathematics is, so to speak, a language in which our intuitive perceptions can be expressed more effectively. 

Now, since physical theories use mathematical structures, their structure should be axiomatic. The axiomatization of physics allows us to apply the deductive method from which theorems and quantitative predictions are derived. Such predictions are usually tested in the light of experiments and when the predictions are corroborated, one says that the model has shown its mettle. On the contrary, if the model is incapable of reproducing the data, it should be discarded. In this sense, the job of a theoretical physicist is to single out the mathematical structures or models that fit the observations.

\subsection{Physical objects}

By analogy with the case of physical assumptions, during the construction of the theory, whether physical objects really exist or not turns out to be irrelevant (because ontologies could have a metaphysical source). This is in view of the fact that the proposed concepts and objects will acquire their physical meaning once the model is faced with experimental evidence. This could be the case of strings, loops, taquions, axions, etc. In some other cases, the experimental observations mold the shape of the theory as well as the properties of its physical objects. For instance, the conception of electron was figured out from observations on electrolysis which suggested a minimum quantity of charge. Later, electrons were conceived as an intrinsic part of the atom and new physical properties such as spin were assigned. In brief, the notion of a `physical' object strongly depends on the structure of the observations and the theoretical framework where the object is interpreted.

\section{Assumptions and principles in the history of physics}
\subsection{Hidden assumptions}
Since ancient times people have built theories based on principles which were considered to be absolute truths, but as time went by some of them have been proven to be actually false. One particular case is the famous assumption that heavier objects fall faster than lighter ones. This principle was held as a physical law for hundreds of years but, as more theoretical and experimental evidence accumulated, Galileo showed that it could no longer be held. In some other cases, theories convey some hidden assumptions. For instance, classical mechanics is based on the three laws of motion and some definitions. In addition to these elements, the theory tacitly presupposes some unnoticed or disregarded assumptions such as: (a) measurements do not affect the physical system under study; (b) the speed of propagation of physical entities can have any velocity, even infinite. (c) Physical quantities are continuous. Assumption (a) fails in the microscopic realm. This issue was rectified by quantum mechanics (QM) with the introduction of a powerful postulate: the uncertainty principle. The principle states, among other things, the probabilistic character of a measurement due to the fact that the measuring instrument considerably influences the response of the system under study. Assumption (b), on the other hand, finds its restriction within the context of relativity theory in which there is a maximum speed for the propagation of physical entities. And finally, assumption (c) also finds limitations in QM where some physical magnitudes are discrete. If we extrapolate this reasoning, we can figure out that surely our modern theories are still incomplete and may need a deep revision.

\subsection{Some physical assumptions in the history of physics}
In view of our previous discussion, it is worth extracting from our theories a short list of some of the most typical assumptions. This will help us to be aware of how physics erects and sculpts its `reality'. It is not the intention of this section to discuss either the truthfulness or the falsehood of each example; for it is evident that some can be true, false or uncertain. As we discussed above, what is of real value for physics is their usefulness in solving the problems that physics has at a given moment of its history. Some of the assumptions are:
\begin{itemize}
  \item  Time flows equally for all observers regardless of their state of motion or its position in a gravitational field.
  \item  The earth/sun is the center of the universe.
  \item Rigid bodies, aether, vacuum and fundamental particles (atoms) exist.
  \item  There is no speed limit for the propagation of physical entities.
 \item  The measuring process does not affect the system under study.
  \item  Space, time, and all other physical magnitudes are continuous/discrete.
 \item  Light is a wave/corpuscle.
 \item Space is static/dynamic; space is a condensed state of matter, space is a network of relationships.
    \item Energy, charge, torque, linear and angular momenta are conserved.
 \item The laws of physics were created along with the creation of the universe and they do not evolve with time.
 \item All particles of a particular class are identical (e.g., electrons, quarks, positrons, etc.).
 \item The principle of: relativity, general covariance, uncertainty, equivalence, causality, exclusion, cosmology, etc. hold.
  \item There is an absolute system of reference.
 \end{itemize}
Some of these assumptions have been definitely discarded, some are still in use and some others have been reconsidered several times in several epochs. From these three cases, we would like to deal with the last one. One of the most famous cases is the reintroduction of the heliocentric model, first proposed by Aristarchus of Samos in the third century B.C. The model remained in the shadow for many centuries until it was revived by Copernicus et al. in the XV century. The history of science tells us that this model was by far more superior in describing the celestial mechanics than the Ptolemaic system. Another famous case is related to the nature of light. In 1905 Einstein reintroduced into physics the almost forgotten notion that light is a particle, just as Newton had put forward more than two centuries ago. Armed with this idea, Einstein built a rational explanation of the photoelectric effect discovered in 1887 by Heinrich Hertz. These two examples teach us that no matter how old or controversial an assumption might be, its potential to solve problems justifies its reestablishment as a scientific hypothesis. 

Unquestionably, some assumptions have caused a great impact more than others, not only to the structure of a given theory but also to the whole evolution of physics. Due to their preeminent influence, this kind of proposals deserve both a special attention and a scrupulous assessment; for their arbitrary rejection could be detrimental for the progress of physics. In what follows, we shall discuss the last assumption from the list above. I shall argue that this is one of the principles that physics should revive if physics wishes to make considerable headway for years to come. To this end, I shall try to dissipate some of the misconceptions that have been appended to it for more than a century.

\section{The Principle of Relativity is not at variance with the Preferred System of Reference}
\subsection{Newton's absolute space}
When Newton developed his laws of motion, he thought that they were valid in absolute space (AS). He contended that the water inside the famous bucket was rotating relative not to the bucket but to AS.\cite{newton1} This experiment gave him confidence that any body possesses not only apparent (or relative) motion but also genuine motion and such motion can only be relative to space itself, or generally speaking, relative to a PSR. From this, it follows that if bodies move relative to AS, then this entity has to be something endowed with physical properties and our disquisition would reduce to identify them. For Newton, AS was an homogenous and isotropic background in which material bodies were embedded, some sort of rigid container mathematically represented by Euclidean space. What we all learn at school, on the other hand, is that this entity is not composed of a material substance, rather, it is total emptiness. It is not clear if Newton agreed with this view, but we have evidence that he actually thought that there was an ethereal and pervading material substance conveying gravitational interactions. In a letter sent to Bentley in 1692, Newton wrote: 
\begin{itemize}
  \item [] \textsf{It is inconceivable that inanimate brute matter should, without the mediation of something else which is not material, operate upon and affect other matter, without mutual contact, as it must do if gravitation in the sense of Epicurus be essential and inherent in it. And this is one reason why I desired you would not ascribe 'innate gravity' to me. That gravity should be innate, inherent, and essential to matter, so that one body may act upon another at a distance, through a vacuum, without the mediation of anything else, by and through which their action and force may be conveyed from one to another, is to me so great an absurdity, that I believe no man who has in philosophical matters a competent faculty of thinking can ever fall into it.}
\end{itemize}

Newton's theory of gravitation assumes that between two celestial bodies there is absolutely nothing mediating the interaction, instead, the alleged interaction occurs by gravitational fields acting at a distance. We note from this letter, however, that his theory does not reflect his actual view. His words seem to imply that he did not even believe in total emptiness, since in Newton's time by the word `vacuum' people understood `devoid of matter'.  In spite of this, what is relevant for us is that he established in his theory that space was immovable. This assumption was precisely what the philosopher Ernst Mach disliked.\cite{mach} For he questioned the scientific utility of an entity that exists but is not affected by the matter it contains. Mach replied to the bucket experiment arguing that the water moved relative not to AS but to the stellar matter surrounding the bucket, because for him only relative motion was possible. Although Mach's argument is weighty, it is not clear what physical substance mediates the gravitational force. It was the insight of  Einstein that shed light on the problem some years later. Starting in 1905, Einstein rejected the \ae ther as the medium for the propagation of light and, by doing this, he left `physical' space again absolutely empty, just as in Newton's theory. Einstein immediately realized this flaw and from 1907 to 1916 he embarked in a historical journey to try to `materialize' Mach's ideas.\cite{einstein2,einstein4,einstein5,einstein6} In the Einsteinian vision of the universe, space is mathematically represented by a pseudo-Riemannian manifold characterized by the metrical field $g_{\mu\nu}$ which contains the gravitational potentials. As a consequence, one has to conclude that the water in the bucket moves relative to the gravitational fields (GF). Einstein then finally replaced the material substance, conceived by both Newton and Maxwell, by the metric field.\cite{einstein10} Since then, the assumption that space can be made up of a material substance has been ruled out from physics (we will return to this topic below). Yet, physics has never ignored the power of intuition. In 1933 the Swiss astronomer Fritz Zwicky discovered some anomalies --- now known as dark matter --- in his studies of the Coma galaxy cluster. This evidence clearly suggests that there is something in space by far more complex than originally thought and that it could be indeed composed of an imponderable and invisible material substance: Newton's substance? The \ae ther that Einstein rejected? Unfortunately, his discovery was ignored for about forty years until Vera Rubin et al. revived it in the 1970s. Still dark matter is one of the most puzzling problems in modern physics.

\subsection{Invariance of Newton's laws}
Let us not digress from Newton's work and bear in mind henceforth the previous discussion. It is well established that Newton's laws are invariant with respect to Galilean transformations. This is in virtue of the Galilean Principle of Relativity (GPR) --- In fact, in The Principia, Newton included this principle as corollary V and justified it with the aid of the second law. It states that all mechanical laws are the same in any inertial system of reference (ISR). But, what is the experimental meaning of this principle? It simply means that no mechanical experiment can tell whether an ISR is at rest or in motion relative to AS (this was well understood by Newton).  The understanding of this statement is vital to make clear that {\it the GPR is not at variance with the existence of the PSR}. This being said, let us consider the following two key questions: 

\begin{enumerate}
\it
  \item Does the fact that the PSR cannot be experimentally detected mean that the PSR does not exist?
  \item If the PSR cannot be experimentally detected, does the assumption become a meaningless assumption?
\end{enumerate}

To grasp the deep significance of these questions, let us contemplate the following situation borrowed from QM and that is beautifully discussed at length by Popper.\cite[pp. 211-217]{popper1} During the development of the atomic model, Niels Bohr imagined that electrons follow orbital paths around the nucleus. He assumed that electrons revolve with a given period $T$ and from this the energy levels $E_n$ and thus the emission spectrum of the hydrogen atom was computed. The key for the success of this approach is the assumption that electron orbits are quantized. However, if we take a closer look at the concept of `path' and scrutinize the principles of QM, we find a serious difficulty. According to Heisenberg's principle, the experimental determination of two correlated observables A and B is limited by the uncertainty relation $\Delta A \Delta B\ge h/4\pi$, where $h$ is Planck constant. In particular, if we assume the observables to be the momentum $p_x$ and the position $x$, the principle tells us that: the higher the precision in the measurement of $x$, the higher the error in the measurement of $p_x$ and vice-versa. This means that it is impossible to experimentally determine the particle's path (i.e., the simultaneous knowledge of both $p_x$ and $x$) with a certainty exceeding the above expression. The reason rests in the fact that the measurement affects the pristine state of the particle. Hence, according to Heisenberg himself, it is meaningless to grant any {\it real} significance to the notion of `path'.\cite{heisenberg30a} As such, the path becomes an unobservable magnitude, i.e., a magnitude unaccessible to experimental verification and therefore it is not useful as a basis for theoretical predictions. The recognition of this, urges us to conclude that measurements cannot serve as a foundation to test physical reality, so to speak, there is no such a thing as physical reality since our instruments are not capable of revealing the true state of a system. In reality, the information we get from our measurements is the outcome of the interaction between the instrument and the system under study. If we conveyed these considerations to the extreme, we would arrive at dramatic conclusions: we would conclude that the factual character of physics is just a chimera. Fortunately, not everything is lost and here the probabilistic and statistical interpretation of QM comes to the rescue. Since we cannot access the precise state of a particle, at least we can tell with some degree of certainty the probable outcome of an experiment.

On the other hand, the formalism of QM through the Sch\"odinger equation allows us to calculate with certainty the particle's path up to a moment before the measurement, i.e., the formalism assumes that there is a path. Evidently, this is at variance with Heisenberg's principle. So, does the path physically exist or not? The answer is not trivial, but being physics a factual science, we understand that the experimental data are essential to sustain the scientific credibility of a theory; for the data collected give us some information of the state of a system. But we have also learned that our measuring processes modify the absolute state. Hence, despite that the path cannot be exactly determined, it is scientifically legitimate to presuppose that the particle's path physically exists, just as the formalism assumes. Admittedly, the fact that the measurement destroys the knowledge of the actual path, does not imply that the physical notion of `path' has no scientific value and, at the same time, does not encourage us to reject QM altogether on the basis that the theory is dealing with unobservables (in the words of Popper, metaphysical constructs). 

One more example of this type is illustrative to reinforce the view that the lack of experimental evidence does not suffice to reject a hypothesis regardless of its apparently unobservable character. Consider the postulate that space and time are continuous. Here once more, we have no conclusive experimental evidence to thoroughly sustain this assumption. In spite of this, good reasons can be advanced for trusting our postulate; actually, our theories have indeed shown that it can be true. Thus, having in mind these two examples, the answer to the first key question is clearly in the negative, for if one accepts the existence of a non PSR, one cannot deny the existence of the PSR since the GPR assures the equality of the mechanical laws in all ISR. Then, the second question is immediately answered also in the negative. From here we conclude that the GPR should not be understood as the exclusion of only one ISR, but quite the contrary, as the inclusion of all of them. Evidently, the arbitrary rejection of the PSR can be, in the long term, detrimental for the advancement of physics because we would deprive our theories from elements indispensable for their logical consistency.

\subsection{Invariance of the laws of physics}
 It is unquestionable that the PSR assumption in classical mechanics resulted highly beneficial for the progress of physics for more than 200 years. The extension of this assumption to electromagnetic phenomena was also very fruitful. It achieved its highest peak with the development of electrodynamics. By the mid 1860's Maxwell predicted that light was some kind of electromagnetic wave that travels through the \ae ther. Some years later, in 1887-8, Hertz could generate Maxwell's waves, leaving no doubt that Maxwell was in the right way.\cite{hertz} Nonetheless, the mere corroboration of electromagnetic radiation did not suffice to establish the existence of the medium. Maxwell was aware that in his equations the \ae ther did not appear, that is, they conserve the same form whether there was \ae ther or not. Yet, for him, the material substance was indispensable to avoid the action at a distance that dominated gravitational interactions. Besides, all known waves hitherto required a medium to propagate and it was natural to assume that light waves could not be the exception. By the middle of the 1870s,  Maxwell's theory was still under construction and many experiments were in line waiting for a satisfactory explanation. In the following years, he expanded the scope of the theory but, unfortunately, he prematurely died in 1879. During the next decade, a new generation of physicists resumed Maxwell's work. For this reason, Oliver Heaviside, Oliver Lodge and George FitzGerald were called {\it The Maxwellians}. These brilliant scientists shaped Maxwell's theory nearly as we know it today.\cite{ohara1987a,hunt1991a} But in spite of the great advances, they did not solve the \ae ther issue, still, the equations had no explicit link with the \ae ther. Fortunately, both Hertz and Paul Dirac (six decades later) also realized Maxwell's problem and promptly modified the equations.\cite{hertz,dirac1951a,dirac1952a} With the aim of accounting for effects of charged bodies in motion relative to the \ae ther, Hertz replaced the partial time derivatives by total (also known as convective, Lagrangian, material or substantial) time derivatives. At that time, his formulation did not attract much attention because some of the predictions were in disagreement with experiments on insulators. Incidentally, modern investigations have revealed that Hertz' formulation was not incorrect at all and that the observed discrepancies can be attributed to quantum effects.\cite{pinheiro} Indeed, Dirac in 1951 also proposed a new electrodynamics and discussed that quantum principles play an important role for reviving the \ae ther concept and when considering the topic seriously, the \ae ther was crucial to build a satisfactory theory of the electron (now known as quantum electrodynamics). 
 
The problem of the \ae ther was not only theoretical but also experimental. It was imperative to show that the ubiquitous substance was not a mere idea. To prove its existence, physicists engaged in an epic hunt by the end of the XIX century. In 1887, Michelson and Morley carried out their famous interferometric experiment which, according to the beliefs of that time, would tell them whether the PSR existed or not (below we dispel some misconceptions about these kind of experiments). As is well known, the results were negative, and by analogy with the experimental implications of the GPR, later, from 1900 to 1905, Larmor,\cite{larmor1897a,larmor1900a} Lorentz,\cite{lorentz4,lorentz2} and Poincar\'e\cite{poincare2} realized that no electromagnetic experiment can tell whether an ISR is at rest or in motion relative to the \ae ther. Such discovery was called simply the {\it Principle of Relativity} (PR) and it is considered as a generalization of the GPR. Thus, in spite of its undetectability, Larmor, Lorentz and Poincar\'e answered the above key questions in the negative, whilst Einstein held the opposite opinion; he was actually appealing to the principle of parsimony.\cite{einstein2} Since no experiment can tell whether an ISR is at `real rest' or in `real motion', Einstein declared that these statements are meaningless. For him, just as Mach, only relative motion is measurable and hence has real meaning. Nonetheless, if we strictly follow this line of thought, motion would adopt a fictitious character. These theoretical perplexities were apparently `overlooked' by Einstein but exposed years later by H. Ives and G. Stilwell when they experimentally corroborated time dilation\cite{ives,ives2} --- We shall discuss in the following sections the importance of the PSR on this issue.

To comply with the PR, physicists were prompted to construct a new dynamics which is now known as {\it Relativistic Dynamics}.\cite{granek} Both Maxwell's laws and the new kinematical and dynamical laws are said to be Lorentz invariant. The new symmetry assures that not only the form of the laws of physics (LP) but also the values of their constants remain the same in all ISR. This inevitably leads us to ask again:  {\it Is then the PR at variance with the existence of the PSR?} Certainly, the answer goes in the negative\cite{dirac1951a,dirac1952a} for no experiment forces us to reject the PSR.\cite{lorentz2,poincare2, bell1} By analogy with the GPR, the PR should be understood not as the discrimination of the PSR, but quite the opposite, as the inclusion of the PSR for the description of physical phenomena. Within this context, Lorentz invariance experimentally means that any experiment carried out in the PSR will lead to the same LP that can be found in any other ISR. The history of physics tells us however that modern theories have discarded it following Einstein's canon.\cite{einstein2,einstein10} But if one upholds the opinion that the PSR is not an issue of parsimony but of usefulness and logical consistency in the physics, one can claim that the assumption that there is no PSR is fundamentally wrong. Let us make some other considerations to support this claim.

\section{Experimental and theoretical considerations in favor of the PSR}
Immediately after the development of the special theory of relativity (SR) a hot debate not only on the existence of the PSR but also on the constancy of the speed of light set in both on theoretical and experimental grounds. Even today many researchers in the fields of physics and the philosophy of physics have kept alive these topics from an epistemological perspective. Thanks to their perseverance, the good news is that substantial advances have been made in the last decades. Although not widely known, specially among the physics community, now it has been understood, some consequential factors that could be fundamental for the future of physics. 

\subsection{Misinterpretation of Experiments:  The Michelson-Morley Experiment}
In the first place, experimental arguments against the PSR have been misleading since the advent of SR. Interferometric and non-interferometric experiments performed during the XIX and XX centuries have been considered as proofs that the \ae ther does not exist. For example, it is common to find in textbooks statements such as: {\it if the \ae ther existed the Michelson-Morley experiment (MMX) would have shown any variation in the fringe shift $N$ of the interference pattern}. This is evidenlty misleading because, as we discussed above, in virtue of the PR, no electromagnetic experiment can tell about the existence of the PSR. This is quite clear from the PR and it would be worth dissipating the misconceptions around the experiment because most interpretations suffer from the same drawback. 

\begin{figure}[tp]
\begin{center}
\includegraphics[width=4cm]{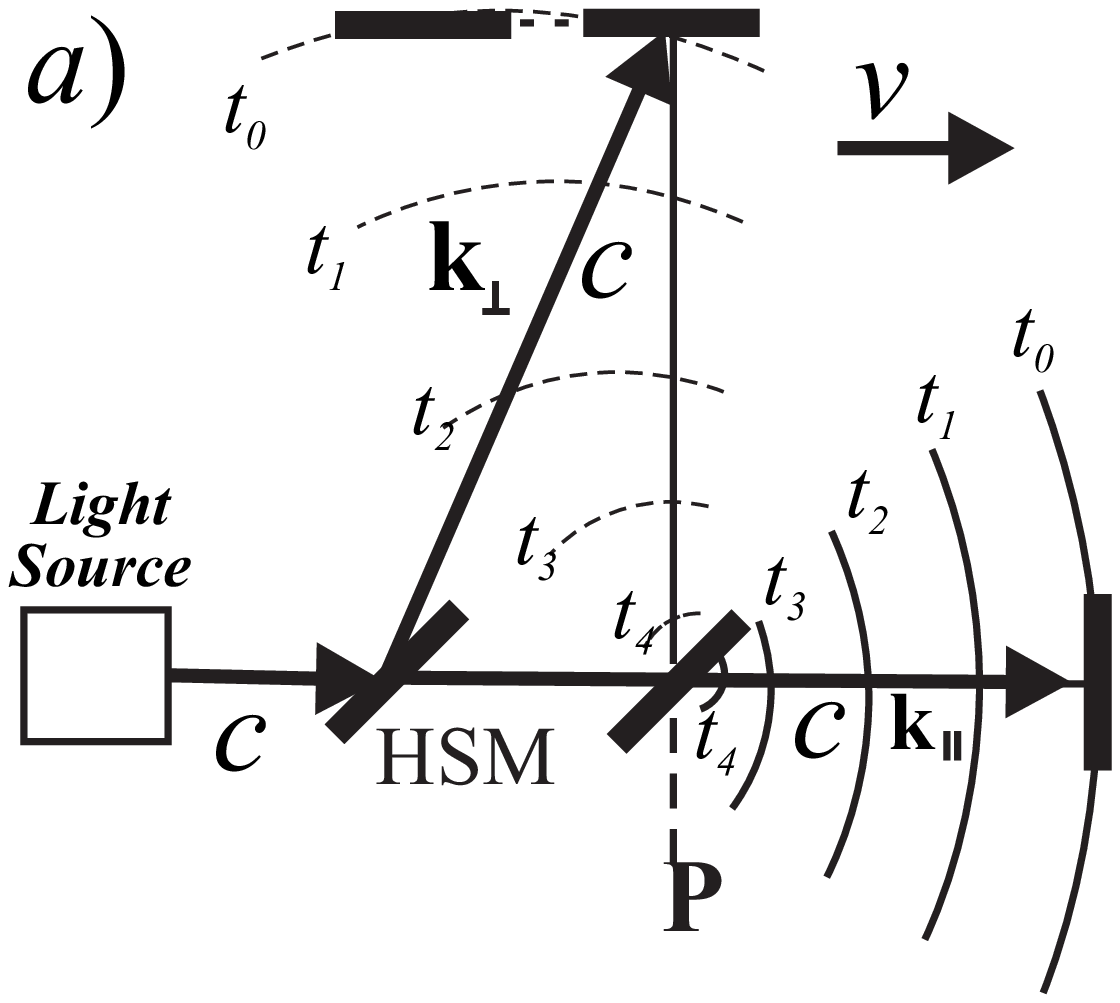} \hspace{0.5cm} \includegraphics[width=4.5cm]{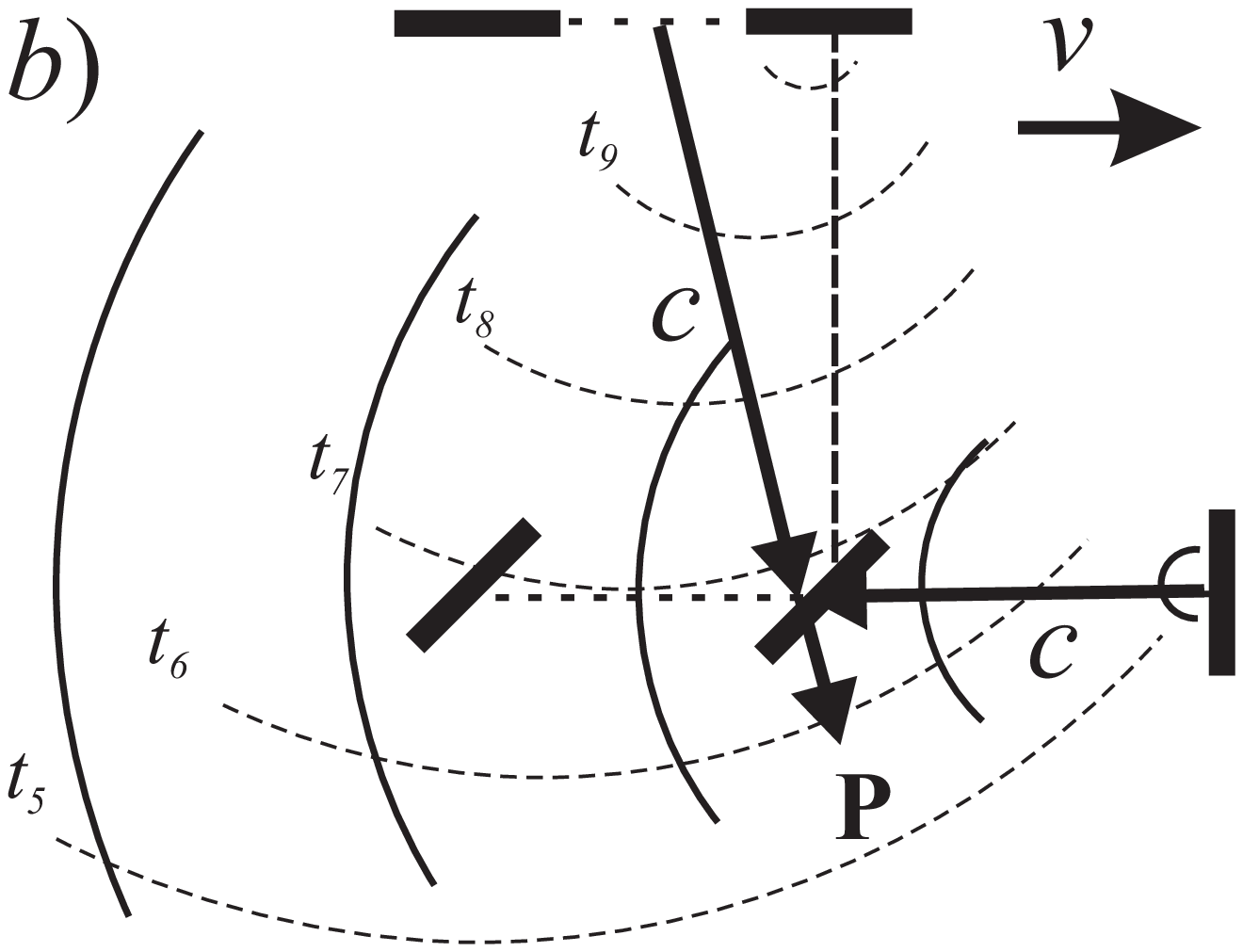}  \hspace{0.5cm} \includegraphics[width=4.5cm]{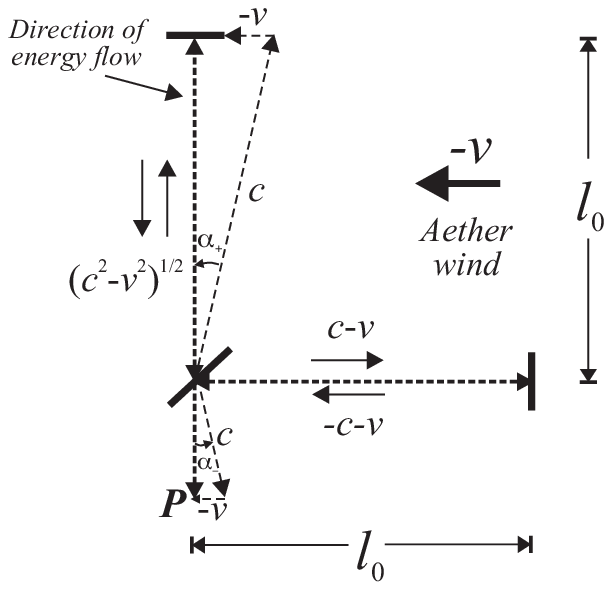}
\caption{The Michelson-Morley experiment. $a$) Forward and $b$) backward advance of light waves as seen from the observer at rest in the vacuum. Arrows represent the propagation vectors of the four wave fronts. Solid arcs are for longitudinal wave motion and dashed arcs for transversal wave motion. c) As judged from Earth, the vacuum is passing by. Using Galilean transformations, the one-way speed of light  becomes anisotropic.}
\label{fisk}
\end{center}
\end{figure}

The experiment had the purpose of measuring the absolute speed of the Earth relative to the \ae ther --- to illustrate my point, I will use the words `vacuum' and `\ae ther' interchangeably. In Fig. \ref{fisk} we show the MMX as seen from both systems of reference, the vacuum ($S$) and the Earth ($S'$) that moves with speed $v<c$ relative to $S$. 
For the sake of illustration, the extension of the waves fronts for the four electromagnetic waves have been accentuated to visualize the Doppler effect and their propagation vectors are displayed as well. According to Michelson and Morley the determination of $v$ would be obtained by measuring changes in the phase difference $\delta$ produced at the point $P$ by the interference of two electromagnetic waves. Then, from the theory of interference as calculated in $S$, the phase difference is
\begin{equation}
\label{eqpha2}
\delta=k(s_{\|}-s_{\bot})=k\delta s=\omega \delta t,
\end{equation}
here $\delta s=c\delta t$ is the difference in the optical path length (OPL) for the waves in the longitudinal and transversal journeys, respectively; and $\delta t=t_{\|}-t_{\bot}$ the corresponding time difference. In these expressions we have used the relations: $c=\omega/k=\nu \lambda$ for the one-way speed of light in vacuum, with $k=2\pi/\lambda$ the wave number, $\lambda$ the wavelength of light and $\omega=2\pi\nu$ the angular frequency. From the preceding formulation it is evident that a fringe shift exists whenever $d\delta/dt\neq 0$. This is achieved in practice by varying the OPL or changing the speed of the beams of light. Recall also that the experiment was designed to revolve, so we have to consider an angular dependence $\theta$ in the phase representing the revolution of the plane of the interferometer with respect to the motion of the Earth (there is still another angle to be considered but to convey our idea we do not need to include it here\cite{articolo,mazur,capria}). Ignoring the length contraction effect, the expression for the phase in the system $S$ is:
\begin{equation}
\label{equ2}
\delta=\frac{\omega}{c}\bigg\{l_0\gamma^2\bigg[(1-\beta^2\sin^2 \theta)^{1/2}- (1-\beta^2\cos^2 \theta)^{1/2}\bigg]\bigg\},
\end{equation}
here $\beta=v/c$, $\gamma=1/\sqrt{1-\beta^2}$, and $l_0$ is the length of the arms as measured at rest in $S$. This equation tells us that $\delta=\delta(\beta,\theta)$, but if the apparatus is not rotated we arrive, to a first approximation, at the traditional expression found in most textbooks: $\delta \approx(l_0\omega/c)\beta^2$. If we assume the Earth as an ISR then $d\delta/dt=0$ and no fringe shift will be observed. So, the rotation of the apparatus is indispensable to observe a fringe shift. The maximum phase occurs after a $\pi/4$ rotation and $N$ is calculated by the difference before and after rotation; then we have $N=\delta_A-\delta_B \approx(2l_0\omega/c)\beta^2$. However, when we consider length contraction in Eq. \eqref{equ2}, we find that the OPL is the same for both light beams, so $\delta=0$ and hence no fringe shift is observed regardless of the variations of $\theta$ and/or $\beta$. This justifies why the experiment failed to observe a positive result. Given this outcome we ask: does this mean that there is no vacuum? To give a definite answer, let us now discuss the physics from the perspective of an observer in $S'$. 

First, we emphasize that in equation \eqref{eqpha2} we have made used of the relation $c=\lambda \nu$ to express $\delta$ in terms of time. This change is possible because in the solutions of the wave equation, $\omega$ and $k$ have a linear dispersion relation, i.e., the group and the phase velocities $V\equiv \partial \omega/\partial k=\omega/k=c$ are the same in all directions (isotropy of the one-way speed of light in vacuum). That the one-way speed of light is isotropic in at least the system $S$, does not follow from SR but it is a direct consequence of electrodynamics. Our problem is then to find out if this is also true in any other ISR in motion relative to the vacuum. Immediately, some closely related questions come to our mind: Where do the alleged anisotropy of the one-way speed of light find in most relativity textbooks come from? What is the physical basis for postulating that `{\it the speed of light is independent of the state of motion of the source or the observer}'? In the last statement it is implied that the velocity of light can depend on the velocity of the source or the observer. But what is the rational source that prompted physicists to conceive such possibility?

Before the discovery of the Lorentz transformations (LT) the only known transformations relating two ISR, moving with relative velocity $\mathbf{v}$, were the Galilean transformations:
\begin{equation}
\label{galtrans}
\mathbf{r'}=\mathbf{r}-\mathbf{v}t; \qquad t'=t.
\end{equation}
Note that the time relation expresses that time flows equally for all inertial observers, meaning that there is a unique rate of flow in all ISR. As a consequence of these transformations, physicists were induced to believe that the speed of light could acquire different numerical values in frames in motion relative to the \ae ther and, in consequence, the wave numbers $k$'s or the frequencies $\omega$'s for each of the light beams involved in an experiment would not take on, in general, the same values. This can be easily shown by applying these transformations to the phase of the wave function $\Psi=a\exp[2\pi i(\mathbf{n}\cdot \mathbf{r}-ct)/\lambda]$, where $a$ is the amplitude and $\mathbf{n}$ a unit vector. After a straightforward calculation, the phase in the system $S'$ becomes: $2\pi i[\mathbf{n}\cdot \mathbf{r}'-(c-\mathbf{n}\cdot \mathbf{v})]/\lambda$. Therefore, in $S'$, the speed of light is $c-\mathbf{n}\cdot \mathbf{v}$, which is anisotropic. Evidently, the error in this prediction is the misapprehension that electrodynamics and Galilean relativity are compatible formulations. This is what Einstein spotted in his famous {\it Gedankenexperiment} about the race with light rays. While Maxwell's theory states that the one-way speed of light is a constant relative to the \ae ther, the Galilean addition of velocities dictates that the speed of light must be velocity dependent. 

Although we have already identified our `na\"ive' mistake, let us further proceed with our analysis. As seen from $S$ (refer back to Fig. \ref{fisk}(c)), the speed of energy flow (or energy flux given by the Poynting vector) for the longitudinal and oblique beams is $c$, therefore the velocity of the energy flow in the $y$-direction is $\sqrt{c^2-v^2}$. According to the observer in $S'$, the vacuum is passing by with velocity $\mathbf{v}'=-v \, \mathbf{\hat{x}}'$, and if we apply Galilean relativity to light propagation, the velocities of the energy flow for the four beams in the frame $S'$ must take on the values:
\begin{equation}
\label{trans}
\mathbf{c}'_{\| \pm} =\pm(c\mp v)\; \mathbf{\hat{x}}'; \quad \mathbf{c'_{\bot \pm}}=\pm \sqrt{c^2-v^2} \;\mathbf{\hat{y}}',
\end{equation}
where $\pm$ stands for forward and backward directions, respectively. Thus, for the parallel direction the wave fronts travel the OPL: $s'_1=l'_0=t'_{1}(c- v)$ and $s'_2=l'_0=t'_{2}(c+v)$; for the forward and backward journeys, respectively. Accordingly, the time spent in the longitudinal journey is $t'_{\|}=(2l'_0/c)\gamma^2$. The time for the transversal journey is calculated as follows. The transversal distance in one direction is $s'_{3}=s'_{4}=l'_0=c'_{\bot}t'_{\bot}/2$. Solving for $t'_{\bot}$ and taking the difference $\delta t'$, we obtain to a first approximation: $\delta \approx \omega(2l'_0/c)\beta^2$. Since the Earth is in motion relative to the vacuum, then $t'=t\gamma^{-1} $ and $l'_0=l_0\gamma^{-1}$. Taking into account these effects in our previous calculations, we also find that $\delta=0$. Showing once more that the experiment cannot determine the Earth's velocity relative to the vacuum.

We must remark, that the absolute speed of light waves never changes regardless of the speed $S'$ relative to $S$, because the light waves travel through the vacuum and its speed is determined by the properties of the medium. If we assume that the medium is static, isotropic, homogeneous and its temperature remains constant, we have no reason to believe that the speed of light would change. The alleged anisotropy of the speed of light is just a {\it fictitious} effect caused by the relative motion between the Earh and the vacuum. This automatically means that the speed of the waves is independent of the motion of the source or the observer (second postulate of SR). Thus, the null result of these kind of experiments does not prove that there is no medium. Some experiments that concluded that there is no medium, made the same `mistake' of convoluting Galilean relativity and electrodynamics.\cite[pp. 518-524]{jackson1999a} 

If we have made clear that no experiment of this kind rules out the PSR, we are faced again with the two key questions above and, therefore, the issue may become only a matter of usefulness and coherence in the logic of a theory. Einstein rejected the \ae ther, first, because, from the theoretical viewpoint, SR could not make special distinctions among ISR; actually, for him the \ae ther assumption was not wrong but appeared to be superfluous. And second, because, from the experimental viewpoint, there was no unambiguous evidence of its existence. Nevertheless, according to the discussion of the previous section, the first argument is weak, for if one follows such line of thought then Newton's AS would have been rejected as well from classical mechanics since the GPR guarantees that all ISR are equivalent. In Einstein's epoch, the second argument had a great weight, however, the discussion given above and the experimental evidence accumulated after the 1930s, strongly disagrees with Einstein's view. The experimental evidence we are referring to is this. Consider the following hypothetical situation. Imagine that before the discovery of relativity, particle accelerators had been already developed. And assume that the ALICE, ATLAS and CMS collaborations at the large hadron collider had released the news, well-known today, that the quantum vacuum is actually a perfect fluid. \cite{nagle} If this fluid were assumed to be at rest and not significantly affected by the presence of material particles it would immediately be identified as the \ae ther or AS; just in the same way as in 2012 many physicists sympathized with the discovered boson at the LHC and identified it as the Higgs boson despite that they did not know yet its other physical properties (spin, etc.). So, if by 1905, physicists had already discovered the presence of dark matter, the background radiation, the presence of a perfect fluid and the Casimir effect, would physicists, despite the success of relativity, have good reasons to discard the medium for light and thus the PSR? Indeed, the answer would be in the negative. The concept would be maintained because the experimental evidence would have suggested its presence. 

\subsection{On the experimental `corroboration' of the second postulate of Special Relativity}
In second place, research on experimental methods used for the measurement of the speed of any physical entity, shows that when the paths of the physical entities involved in the measurement form a closed circuit, what the experiment measures is an average speed, i.e., a harmonic mean of the speed or the so-called two-way speed. This implies that it is not feasible to measure the one-way speed of light.\cite{perez2} Since the second postulate of SR tacitly states that there is a finite isotropic speed $c$, the studies reveal that this postulate has never been experimentally corroborated. Under such scenario, one can raise sharp objections against either SR or electrodynamics similar to those raised against the PSR. But despite that nature conspires to hide from us this crucial knowledge, there are enough reasons to hold the postulate that the one-way speed of light is actually isotropic in vacuum. Once more, this lack of experimental evidence does not impel us to reject our postulate.

\subsection{Misinterpretation of Newton's theory}
The structure of a physical theory is fixed and cannot be modified at will. A genuine scientific theory cannot, when faced with an irreconcilable fact, be rectified a little to make it agree. There are some colleagues (see for instance\cite[p. 6]{landau}) who have claimed that Newtonian mechanics is just as relativistic as SR under the argument that Galilean transformations leave invariant Newton's laws. They even contend, led by the relativity school, that Galilean invariance demonstrates that there is no PSR. This is, of course, nonsense for AS is a principle of the theory and it cannot be arbitrarily eliminated. In fact, we shall see below that Minkowski space plays the role of absolute space in SR.

\subsection{Relative motion leads to quandaries}
\label{relquand}
It is worth discussing briefly how quandaries arise in SR due to the idea that only relative motion is significant. Here I shall consider the case of time dilation which is usually confused with the so-called clock paradox. I do not treat the original clock paradox\cite{einstein2} since it has shown to be misleading.\cite{grunbaum1954} Instead, I slightly modify the situation to identify where the perplexing part of SR is. The problem is related to the topic of relative motion versus absolute motion. Imagine three synchronized clocks placed in line at three equidistant points A, B, and C. Consider that clocks at A and C are moved simultaneously (that is, at $t_A=t_B=t_C=0$) towards B with the same constant velocity $v$ (and by symmetry, the same initial acceleration if you wish). Now we wonder whether time really dilates or not for clocks in motion. (i) According to SR, an observer at rest next to clock B will figure out that, since both clocks A and C are moving towards B at the same speed, they will arrive at B synchronized among each other but lagging behind clock B by the factor $\sqrt{1-(v/c)^2}$. So far so good, but this is not the end of the story. Relative motion strictly dictates that the two clocks A and C are not only moving relative to each other at constant speed $V$, but also relative to the clock B at speed $v$. Since according to Einstein there is no PSR, this means that absolute motion is meaningless and consequently only relative motion is significant. On the basis of this theoretical restriction, it is equally legitimate to judge the situation from the standpoint of an observer in the ISR of clock A. (ii) From this perspective, clocks B and C are approaching clock A at speeds $v$ and $V$, respectively. And by symmetry, when the three clocks meet, the observer at A will find that, both clocks B and C, will lag behind clock A in proportion to their relative velocities, $\sqrt{1-(v/c)^2}$ and $\sqrt{1-(V/c)^2}$, respectively. Moreover, since $V>v$, he will assert that clock C will be lagging behind clock B. (iii) With the same right and by the same argument, a third observer in the ISR of clock C will claim that when the three clocks meet, clocks B and A will be lagging behind clock C in proportion to their relative velocities, etc. Certainly, according to the view that only relative motion is meaningful, the tree options are equally legitimate, although it is obvious that if the experiment is performed the three options cannot be true. In view of these baffling conclusions, it is impossible to decide solely on the grounds of the principles of the theory itself, what would be the {\it actual} outcome of an experiment like this (a similar situation occurs with the stellar aberration). By 1937, Ives and Stilwell realized about these quandaries and discussed the topic at length.\cite{ives, ives2,ives1941a} They carried out a series of experiments to test time dilation and pointed out that the source of the problem is the omission of the \ae ther. If we reintroduce the PSR in our picture, we will have a logical criterion to decide what will be the actual outcome of the experiment since, in this case, only absolute motion is meaningful (below we discuss how to distinguish absolute from relative motion). Even if we were not able to determine the real state of motion of an ISR, we can still theoretically assume that either the ISR of clock B is at rest or moving at speed $w$ relative to the PSR. Under this scenario, we realize that the flow of time of clock B will remain constant at all times whereas the flow of time for the clocks A and C will be altered since they are absolutely moving (for detail calculations on this view see Ives and Stilwell works\cite{ives, ives2,ives1941a}). Therefore, from the absolute point of view, options (ii) and (iii) are na\"ive and can be discarded at once. We are left then with option (i). Whether this option is true or not would depend on the adopted clock synchronization convention, topic which is outside the scope of present work.\cite{perez2} This example constitutes a logical justification to reconsider the PSR. Einstein rejected it because he considered it superfluous, now we see that parsimony leads to logical predicaments. What we learn here is that parsimony is not always the best choice; for if a theory A, assuming the PSR, explains the same amount of observations as another theory B, in which no PSR is assumed, one should chose theory A because it is free from perplexities. The theory A we refer to is not SR with a PSR, but Lorentz'  \ae ther theory \cite{lorentz4} [not to be confused with FitzGerald-Lorentz hypothesis about length contraction].

\subsection{The law of inertia and the conservations laws}
 When we work within the context of Newtonian mechanics, we are usually unaware how the law of inertia was defined.  The law states that: {\it Every body persists in its state of rest or of uniform motion in a right line, unless it is compelled to change that state by forces impressed thereon}. But with respect to what system of reference is this true? To answer, let us imagine that we have two systems of reference, $S$ and $S'$, with $S'$ moving along the $x$ direction with velocity $V$ and acceleration $W$ relative to $S$. We now have a particle that moves also along the $x$ direction with velocity $v$ and acceleration $w$ relative to $S$ as well. According to Galilean relativity, the velocity $v'$ and the acceleration $w'$ in $S'$ are given by: $v'=v-V$, $w'=w-W$. If the particle moves by inertia relative to $S$, we have in $S'$ that $w'=-W\neq0$, that is, the observer in $S'$ cannot  figure out that the particle is moving by inertia. For this reason the law of inertia loses its meaning if we do not specify to what system of reference the law refers to. In consequence, in order for this law to be meaningful, we need to define a special system where the law of inertia holds. Following Newton, such system is AS and by virtue of the Galilean transformations, the law of inertia is also true in any other system moving uniformly relatively to AS. It follows that a system in which Newton's laws hold is an ISR.

The recognition of Newton's laws and AS invites us to accept their consequences since space (Euclidean) and time are isotropic and homogeneous. From these properties, as we know from Noether's first theorem, the conservation of momentum and energy follow. Because Minkowski space-time has these properties, the law of inertia and the conservation of energy-momentum hold also true in SR. In his article on cosmological considerations,\cite{einstein12} Einstein, led by Mach, manifested his disagreement with Newton enunciating the so-called `relativity of inertia': {\it In a consistent theory of relativity there can be no inertia relatively to `space', but only inertia of masses relatively to one another}. This statement is just another version of {\it Mach's principle}. To incorporate this principle in General Relativity (GR), the law of inertia and the sacred conservations laws had to be redefined. Lorentz, Klein, Einstein, et al. soon realized this `inconvenience'  and tried to amend it. The reason is quite evident: for Einstein, inertia is due to the presence of other masses and relativistic space is, in general, dynamic (non-Euclidean). In an attempt to save energy conservation, Einstein introduced the pseudo-tensor $t_i^k$ and claimed that the total momentum and energy, $J_i=\int(T_{\mu\nu}+t_i^4)dV$ (with $T_{\mu\nu}$ the stress-energy tensor), of the closed system are, to a large extent, independent of the coordinate system.\cite{einstein16, pauli57a} The contemporary version acknowledges, however, that the `conserved quantities' are not in general conserved in GR and other diffeomorphism covariant theories of gravity.\cite{wald00a,szabados09a} The problem consists in that in GR, gravitation is represented by the metric tensor $g_{\mu\nu}$ (that underlies the geometry of space) and the gravitational energy contained in the geometrical part cannot be, in practice, localized. In mathematical terms, this is implied in the divergence of the stress-energy tensor
\begin{equation}
\label{econserv}
T^{\mu\nu};_\mu=0,
\end{equation} 
which expresses the exchange of energy-momentum between matter and the gravitational field. For asymptotically flat and stationary spacetimes at infinity (i.e., spaces that tend to Minkowski space), one can always find an energy conservation law by integration of Eq. \eqref{econserv}. But, it is no longer possible for general spacetimes. 

As for the law of inertia, Einstein worked out a cosmological model where he first considered the scenario of an open and expanding universe. \cite{einstein12} To solve the gravitational field equations, one needs to provide the boundary conditions. By a suitable choice of a reference system, the $g_{\mu\nu}$ in spatial infinity tends to Minkowski metric $\eta_{\mu\nu}$. He rejected this possibility because he first realized that the reference system would represent a PSR contradicting the `{\it spirit of relativity}'; and, secondly, because this choice would discriminate Mach's principle. He then opted for avoiding boundary conditions at infinity and considered the possibility of a finite and closed universe. For this purpose, he modified his field equations introducing the famous cosmological constant $\Lambda$ (for details on the physical meaning of $\Lambda$, see A. Harvey et al.\cite{harvey00a}). The modified equations read:
\begin{equation}
\label{eineq}
R_{\mu\nu}-\frac{1}{2}Rg_{\mu\nu}=T_{\mu\nu}+\Lambda g_{\mu\nu},
\end{equation}
where $R_{\mu\nu}$ is the Ricci tensor and $R$ the scalar curvature. With this new term, he thought he had succeeded not only in `satisfying' Mach's principle but also in removing the boundary conditions. His joy, however, did not last much because, in the same year, the Dutch astronomer de Sitter found a vacuum solution in which matter ($T_{\mu\nu}=0$) was not necessary to define an ISR.\cite{weinberg2} Three decades later, Pauli recognized that the problem was still open\cite[p. 182]{pauli57a} and Steven Weinberg expressed in his book of 1972 that the answer given by GR to this problem through the equivalence principle `{\it lies somewhere between that of Newton and Mach}'.\cite[pp. 86-88]{weinberg72a} More recently, some physicists claim that the Lense-Thirring effect contemplates some effects of Mach's principle,\cite{bondi97a} although, most specialists agree that GR is neither completely Machian nor absolutely relativistic,\cite[p. 106]{barbour95a} implying that, after all, both the PSR and Newton's law of inertia are still very alive. In addition to this failure, a closer inspection of the ontology of space in GR reveals peculiarities that require a careful examination. 

\subsection{Are space and vacuum the same physical entities?}
The understanding of the nature of the vacuum might be crucial to achieve the TOE because this problem is closely related to the energy density of the vacuum $\rho_{vac}$ and the cosmological constant problem. The `na\"ive' calculations for $\rho_{vac}$ obtained from quantum field theory (QFT), yield a value that differs by about 120 orders of magnitude when compared to the value $\rho_{\Lambda}$ obtained from cosmological observations where it is believed that GR is the correct theory. Up to now, physicists are still perplexed for such a big difference.\cite{weinberg2,carroll01a,Padmanabhan03a} This can indicate that either GR, or QFT, or both are in need of serious revision. After many years of study, some researchers suspect (as I do) that the geometrical representation of space may not be the best choice for the future of physics.\cite{ontospace1} In fact, this notion seems to be at variance with the notion of vacuum in QFT. In this theory the vacuum has a ground state energy different from zero, the so-called zero point energy. Nevertheless, it appears that GR has distinct understanding. Taking a closer look at Eq. \eqref{eineq}, we see the following peculiarities: The left hand side represents the geometry of space where the energy of the GF is included. On the right hand side, we find the $T_{\mu\nu}$ where we include `matter' and the $\Lambda$-term that can be understood as a repulsive force due to the vacuum energy (known as dark energy). Since the latter can be put on the left hand side too, one can interpret it as gravitational energy rather than vacuum energy (this is still under debate). If we leave it on the right side, the vacuum is viewed, in its rest frame, as a perfect fluid with energy density $\rho_{vac}$ and isotropic pressure $p_{vac}$, both related by $w=-P_{vac}/\rho_{vac}$. For this fluid the stress-energy tensor reads
\begin{equation}
\label{energyvac}
T^{vac}_{\mu\nu}=(\rho_{vac}+p_{vac})u_{\mu}u_\nu+p_{vac}g_{\mu\nu},
\end{equation}
where $u^\mu$ is the fluid four-velocity. If we assume a motionless fluid, the first term in this expression is zero. Hence, $\rho_{vac}$ is equivalent to $\Lambda$ and it is legitimate to assume that $\rho_{vac}=\rho_{\Lambda}=\Lambda/(8\pi G)$. Now, Einstein equations for `empty space' with no vacuum energy ($\Lambda=0$) read $R_{\mu\nu}=0$. Solutions of these equations are Minkowski and Euclidean spaces. It is worth noting that Einstein considered Minkowski metric as a special case of metric with constant gravitational potentials $\Phi$. Since $\Phi=$ constant, there is no GF in this space. Moreover, from the geodesic equation 
\begin{equation}
\label{geoeq}
\frac{d^2x^{\mu}}{ds^2}=-\Gamma^\mu_{\alpha\beta}\frac{dx^{\alpha}}{ds}\frac{dx^{\beta}}{ds},
\end{equation}
where $\Gamma^\mu_{\alpha\beta}$ is the Christofel symbol and $s$ a scalar parameter of motion, it follows that a test particle moves in straight line (Newton's law of inertia). This rationale also applies to Euclidean space, implying also that, in this space, there is a constant gravitational potential and, therefore, zero GF. Nonetheless, it is difficult to reconcile ourselves with this interpretation given that there are no sources of gravitation and given that, in Newtonian mechanics, Euclidean space represents AS, i.e., total emptiness. We are thus tempted to think that if Euclidean space represents the PRS in Newton's theory, Minkowski space represents the PSR in SR. Naturally, for GR, Minkowski space is not a realistic space, although our analysis is exposing the substantival character of space in GR. Considering now that the vacuum has nonzero energy, Einstein's equations read $R_{\mu\nu}-\frac{1}{2}Rg_{\mu\nu}=\Lambda g_{\mu\nu}$. If $\Lambda>0$, $\rho_{vac}>0$ and $p_{vac}<0$, one of the solutions is the {\it de Sitter flat space}. If, $\Lambda<0$, $\rho_{vac}<0$ and $p_{vac}>0$ we have the {\it anti-de Sitter space}. In the former, space is open and expands; and in the latter, space is close and the expansion decelerates. So, tests particles will move apart in a de Sitter space; implying inertia without matter (in contradiction to Mach's principle). Finally, if $T_{\mu\nu}\neq 0$, the field equations are of the form \eqref{eineq} and one of the solutions is the well-known Friedmann-Walker-Robertson space which represents our `real' expanding universe. In any case, we see that regardless of our considerations on the right hand side of Eq. \eqref{eineq}, there is always space (except when all components of $g_{\mu\nu}$ are zero) which is subsequently filled, according to our considerations, with `stuff'. In this sense, GR represents `space' as a container (the pseudo-Riemannian manifold) that responds according to the energy-matter content. Is there any problem with this? Indeed, in GR we can have space seen as perfect emptiness, and space seen as an energetic perfect fluid (the vacuum). We see that, just as the Euclidean manifold plays the role of the background in Newtonian mechanics, the pseudo-Riemannian manifold along with the metric is playing the role of a substratum, since space can exist even if $T_{\mu\nu}=0$ and $\Lambda=0$. 

The fact that in GR is possible to have space without sources of any kind seems to be in contradiction with the notion of vacuum as seen from QM and electrodynamics. Whilst geometrical spaces have no electromagnetic properties per se (compare to the Reissner-Nordstr\"om metric), the vacuum of electrodynamics has intrinsic finite electric permittivity $\epsilon$ and magnetic permeability $\mu$. The assumption of the existence of a perfectly empty space is, just as the assumption of the existence of rigid bodies, a false but useful assumption. That the vacuum is an actual physical entity can be demonstrated even from the perspective of electrodynamics. To get the feeling of this, consider the following situation.\cite{urban13a} Suppose that a coil with $n$ turns is energized and carries a current $I$. Accordingly, the magnetic induction of the coil is $B=\mu_0nI+\mu_0M$, where $nI$ is the magnetic intensity and $M$ is the magnetization induced in the coil. If we carry out an experiment where we keep the current constant and reduce the density of matter, $B$ decreases. As we continue to eliminate `all matter' then $M=0$ and $B=\mu_0nI$. This result experimentally demonstrates that the vacuum is a {\it paramagnetic medium} with magnetic permeability $\mu_0=4\pi 10^{-7}$ N/A$^2$. And because this property is exclusive of matter, the experiment tells us that the vacuum is not deprived of `material substance' at all. In contrast, if physical space were totally empty, one would expect null electromagnetic properties. 

On the other hand, the field of condensed matter has made important advances, particularly, in the field of Bose-Einstein condensates and superfluids. Giving the mathematical analogies of these systems with the quantum vacuum, some physicists have suggested that the vacuum could be a condensed state of matter.\cite{volovik} One of the consequences of this approach is that perhaps the equivalence principle and some other symmetries such as Lorentz invariance and gauge invariance are not fundamental at all but emergent from the low-energy sector of the quantum vacuum. Indeed, F. Witenberg showed that assuming the vacuum as a kind of plasma composed of positive and negative massive particles interacting by the Planck force over a Planck length, one can derive QM and Lorentz invariance as asymptotic approximations for energies small compared to the Planck energy.\cite{winterberg07a} He finally concluded that Minkowski spacetime is superfluous for physics provided that Lorentz invariance is assumed as a dynamic symmetry, just as conceived in Lorentz' \ae ther theory where length contraction and time dilation are explained in terms of atomic deformations and motions through the medium. Certainly, this would imply that electromagnetic fields, and no less particles, are states and excitations, respectively, of the vacuum. Following a similar line of thought, M. Urban et al. recently showed that the origin of the speed of light (and the permeability $\mu_0$ and permittivity $\epsilon_0$ constants) is the result of interaction of photons with fermions pairs spontaneously produced in the quantum vacuum. This implies, again, that the vacuum is the medium for light and that the speed of light is strictly defined relative to it. As for the law of inertia, B. Haisch et al. put forward a quantum mechanism to justify its origin. They showed that inertia can originate from the quantum vacuum without alluding to Mach's principle.\cite{haisch01a} Admittedly, all this evidence strongly suggests that the vacuum is some sort of diluted material fluid. If we trust this view, the energy density $\rho_{vac}$ found from QFT might be downright correct. Admitting the vacuum as a material substance capable of transmitting gravitation, just as Newton devised it, prompts us to deeply revise the geometrical interpretation of space and gravitation in GR.

\section{General relativity is not fully relativistic and the speed of light is not constant}
\subsection{Absolute motion versus relative motion}
In his celebrated scholium,\cite{newton1} Newton taught us how to distinguish false motion from real motion; there he wrote:
\begin{itemize}
  \item [] 
   \textsf{ The causes by which true and relative motions are distinguished, one from the other, are the forces impressed upon bodies to generate motion. True motion is neither generated nor altered, but by some force impressed upon the body moved; but relative motion may be generated or altered without any force impressed upon the body...}
\end{itemize}
Then, at the end, he topped off:
\begin{itemize}
  \item []   \textsf{But how we are to collect the true motions from their causes, effects, and apparent differences; and viceversa, how from the motions, either true or apparent, we may come to the knowledge of their causes and effects, shall be explained more at large in the following tract. For to this end it was that I composed it.}
\end{itemize}
Newton's Principia was completely devoted to demonstrate that absolute motion exists and can be distinguished from relative one by forces. 
In Newton's theory, an observer in a non-inertial system (NIS), say the Earth, that rotates with angular velocity $\mathbf{\omega}$ and translates with acceleration $\mathbf{\ddot{R}}$ relative to an ISR will observe a series of forces acting on a particle of mass $m$: 
\begin{equation}
\label{ftrans}
\mathbf{F}'=\mathbf{S}+m\mathbf{g}-m\ddot{\mathbf{R}}-m\dot{\omega}\times \mathbf{r}-m \mathbf{\omega}\times (\mathbf{\omega}\times \mathbf{r})-2m\mathbf{\omega}\times\mathbf{v},
\end{equation}
where $\mathbf{S}$ is the sum of external forces, $\mathbf{g}$ the gravitational acceleration and $\mathbf{v}$ the particle's velocity as measured on Earth. The fourth term appears in case $\omega$ is not constant, the fifth term is the centrifugal force and the last one is the Coriolis force. All these additional forces are named {\it inertial, fictitious or pseudo forces}. The adjective `fictitious'  and the prefix `pseudo' speak for themselves. In this theory, {\it these are not real forces} because their nature arise from relative motion. Not convinced, a relativist will claim that the Earth observer `feels', i.e., measures these forces and, hence, they are real for him; consequently, the adjective `fictitious' fades away. A Newtonian in turn will reply that if motion were purely relative, the Earth could be considered as static frame subjected to the pseudo-forces of Eq. \eqref{ftrans} and, in consequence, the view that the world rotates around the Earth would be equally true. This line of reasoning will send us back to the idea of the Earth as the center of the universe and one would not be able to decide whether the Earth {\it really} rotates or not (similar to the time dilation quandary discussed in section \ref{relquand}). For a Newtonian, the relativist view is, needless to say, na\"ive. For if an experiment could be conceived to measure the effects of the pseudo-forces, we would be demonstrating that AS exists. The Focault pendulum is a beautiful example that Newton was right. Exploiting the effects of the Coriolis force, the experiment not only gives geocentrism a {\it coup de gr\^ace}, but also informs us that absolute rotation can be measured even if we were enclosed in a laboratory without observing the fixed stars. The experiment shows that the Earth's angular velocity relative to AS (Euclidean space) can be determined by just measuring the rotation of the oscillation plane as function of time. Likewise, the Michelson-Gale experiment shows clear evidence that, without looking at the sky, the Earth absolutely rotates relative to the vacuum.\cite{michelson25a} This experiment not only measures $\omega$ but also teaches us that the vacuum is the medium for light. If we now judge these experiments from the standpoint of SR, the Earth revolves relative to a system either in motion or at rest relative to the Minkowskian background (physically speaking the vacuum). If $\omega$ is small, the calculations from both SR and Newton's theory agree. And what does GR have to say about this? For GR, as in the case of the Newton's bucket, the Earth rotates relative to its GF so that the fictitious forces become genuine GF (see the Kerr field and the Lense-Thirring effect). In the case of the Focualt experiment, GR includes tiny corrections that, in practice, cannot be distinguished from Newton's results. We thus see once more that $g_{\mu\nu}$ plays the role of background for the rotation of the Earth, by analogy with the Euclidean metric in Newtonian theory. But just as one cannot place a system of reference at absolute rest relative to AS, one cannot place a system of reference at rest relative to the GF. Thus, to determine $\omega$ astronomers use Eq. \eqref{trans} and assume a special ISR, the so-called fixed-space system or International Celestial Reference System. Such system, evidently, is an ideal candidate for a PSR.

In the development of GR, Einstein sought to justify inertia, and therefore rotational motion, relative to the masses of the universe through both Mach's principle and the equivalence principle. We saw above that he did not succeed. Furthermore, Einstein did not succeed either in creating a fully relativistic theory. This means that not all systems of reference are equivalent. That this is the case can be seen from the principle of general covariance.\cite[pp. 91-93]{weinberg72a} In 1917, E. Kretschmann\cite{kretschmann17} recognized in a critical study of GR, that the principle does not imply that the LP most be relativistic, but only that their form must be the same under general coordinate transformations. In fact, even Newton's laws can be written in covariant form. Thus, general covariance is not a PR but a principle that imposes restrictions between matter and geometry;\cite{norton93,dieks06a} for this reason, John Wheeler suggested that Einstein's theory should be called, instead of `general' relativity, {\it Geometrodynamics}. Today, some physicists still look for a fully relativistic theory where Mach's principle could be embraced.\cite{barbour95a} This, indeed, indicates that GR is not hermetic to accept a PSR, even going against its own spirit. Both astronomy and cosmology have always been in need of a special system of reference to assess the celestial dynamics and define a cosmic time. The cosmic microwave background radiation also strongly demands a special system. It seems to me that the PSR is valuable to satisfactorily account for physical phenomena at all scales.\cite{consoli04a,jacobson01a,scarani00a} 

\subsection{Covariance and the variation of the speed of light}
\label{covalight}
Before we close this treatise, it is worth elucidating the fact that SR has actually only one postulate, i.e., the PR, since the second one is already tacitly included in electrodynamics. This postulate is valid insofar as one deals with ISR, but invariance no longer holds for NIS --- or appealing to the equivalence principle for systems of reference in GF ---. This means that the value of their constants and physical quantities may acquire different values in different NIS. As we showed in Eq. \eqref{ftrans}, the same occurs in Newton's theory. Covariance, by contrast, only demands that the form of the LP must remain the same. As early as 1911, Einstein was aware of this.\cite{einstein4} He knew, for instance, that the only cause that could change the path of light is by varying the speed of the different parts of a wave front. During the development of GR, he emphasized that the assumption of the constancy of the speed of light must be abandoned for NIS.\cite{einstein4,einstein5,einstein6} However, the principle of general covariance (also known as diffeomorphism covariance or general invariance) demands that the metric tensor $g_{\mu\nu}$ must change whereas all constants must remain the same under general coordinate transformations. Since then it is widely believed that the speed of light is a universal constant at any point of a GF. This could be true insofar as we understand space as GR does, but we have shown above that the vacuum can be seen as a diluted material fluid. Under this assumption, we can reinterpret the bending of light just as a simple refraction phenomenon. One can keep the vacuum static and assume it as a inhomogenous medium with degraded refraction index that vary as function of position in the GF. The gradient depends on the gravitational potential which, in turn, will make the speed of light function of position. Thus, within this context, the `warping' of space can be physically understood as the change in the density of the medium.\cite{hao} Certainly, this will not account for the perihelion of Mercury or other gravitational phenomena, but it gives us a hint on how to build a unified theory and reinterpret gravitational effects. 

\section{Final remarks}
Throughout the course of this treatise I briefly reviewed the role played by the PSR in physics. In doing so, I presented a series of epistemological, experimental and theoretical arguments to dispel the series of misconceptions around this central tenet, and, at the same time, I gave good reasons to champion its reintroduction into physics. I also pointed out that the geometrization of space may not be the most appropriate for the future of physics. Instead, the experimental evidence at hand suggests that space is a dynamical condensed state of matter. Due to the lack of space, I cannot discuss here the progress that has already been advanced based on these radical ideas and I prefer to leave it for a future contribution. The purpose of this work is to show that the PSR is not in conflict with physics and that the vacuum can be conceived in a different way. Once we accept this, the next step is to unify the concepts of particle and wave using the notion of quasiparticles. In this sense, a field would become a state of the vacuum and a particle an excitation. The implications of this insight may impact physics at all scales leading to the TOE without invoking exotic assumptions (multiverses, extradimensions, etc.). In my opinion, there are enough experimental and theoretical elements for a new revolution in physics. Thomas Kuhn taught us that a paradigm shift might be a thorny episode in the evolution of science.\cite{kuhn} The PSR assumption constitutes a paradigm shift that would request a drastic change in the way of understanding reality. Some `established' facts such as the expansion of the universe and the big bang model may need to be revised in the light of this new paradigm. Inevitably, this will lead us at some point to the bucket problem. And just as Newton held, here it is claimed that the water moves relative to the vacuum, provided that we understand elementary particles as quasiparticles and the vacuum as a dynamical `material' fluid.

\section*{Acknowledgements}
 This research was supported by the Natural Sciences and Engineering Research Council of Canada, the Canada Research Program and CONACYT Mexico under grant 186142. The author is thankful to the University of Saskatchewan, Prof. Alex Moewes for his support in this project, and the FQXi organizers of the 2012 contest for opening the doors to new and fresh ideas fundamental for the progress of physics.

\bibliographystyle{unsrt}
\bibliography{pfrrref}

\end{document}